\documentclass[sigconf,natbib=true,anonymous=false]{acmart}

\AtBeginDocument{%
  \providecommand\BibTeX{{%
    \normalfont B\kern-0.5em{\scshape i\kern-0.25em b}\kern-0.8em\TeX}}}

\copyrightyear{2023} 
\acmYear{2023} 
\setcopyright{acmlicensed}\acmConference[SIGIR '23]{Proceedings of the 46th International ACM SIGIR Conference on Research and Development in Information Retrieval}{July 23--27, 2023}{Taipei, Taiwan}
\acmBooktitle{Proceedings of the 46th International ACM SIGIR Conference on Research and Development in Information Retrieval (SIGIR '23), July 23--27, 2023, Taipei, Taiwan}
\acmPrice{15.00}
\acmDOI{10.1145/3539618.3591626}
\acmISBN{978-1-4503-9408-6/23/07}

\usepackage{xspace}
\usepackage{multirow, diagbox}
\usepackage{pifont}
\usepackage{enumitem}

\newcommand{\xmark}{\text{\ding{55}}}
\newcommand{\ymark}{\text{\ding{52}}}
\newcommand{\framework}{UIA\xspace}

\acmConference[Conference acronym 'XX]{Make sure to enter the correct
  conference title from your rights confirmation emai}{June 03--05,
  2018}{Woodstock, NY}
\acmPrice{15.00}
\acmISBN{978-1-4503-XXXX-X/18/06}

\makeatletter
\def\old@comma{,}
\catcode`\,=13
\def,{%
  \ifmmode%
    \old@comma\discretionary{}{}{}%
  \else%
    \old@comma%
  \fi%
}
\makeatother

\begin{document}


\title[A Personalized Dense Retrieval Framework for Unified Information Access]{A Personalized Dense Retrieval Framework for\\ Unified Information Access}

\author{Hansi Zeng}\authornote{A part of this work was done while Hansi Zeng was an intern at Lowe's.}
\affiliation{%
  \institution{University of Massachusetts Amherst}
  \country{USA}
}
\email{hzeng@cs.umass.edu}

\author{Surya Kallumadi}
\affiliation{%
  \institution{Lowe's Companies, Inc.}
  \country{USA}
}
\email{surya@ksu.edu}

\author{Zaid	Alibadi}
\affiliation{
    \institution{Lowe's Companies, Inc.}
    \country{USA}
}
\email{zalibadi@email.sc.edu}

\author{Rodrigo	Nogueira}
\affiliation{
    \institution{University of Campinas}
    \country{Brazil}
}
\email{rfn@unicamp.br}

\author{Hamed Zamani}
\affiliation{%
  \institution{University of Massachusetts Amherst}
  \country{USA}
}
\email{zamani@cs.umass.edu}

\begin{abstract}
Developing a universal model that can efficiently and effectively respond to a wide range of information access requests---from retrieval to recommendation to question answering---has been a long-lasting goal in the information retrieval community. This paper argues that the flexibility, efficiency, and effectiveness brought by the recent development in dense retrieval and approximate nearest neighbor search have smoothed the path towards achieving this goal. We develop a generic and extensible dense retrieval framework, called \framework, that can handle a wide range of (personalized) information access requests, such as keyword search, query by example, and complementary item recommendation. Our proposed approach extends the capabilities of dense retrieval models for ad-hoc retrieval tasks by incorporating user-specific preferences through the development of a personalized attentive network. This allows for a more tailored and accurate personalized information access experience. Our experiments on real-world e-commerce data suggest the feasibility of developing universal information access models by demonstrating significant improvements even compared to competitive baselines specifically developed for each of these individual information access tasks. This work opens up a number of fundamental research directions for future exploration.
\end{abstract}



\maketitle

\section{Introduction}
\label{sec:intro}
Information access systems, such as search engines and recommender systems, play a key role in the Web ecosystem. Often, a combination of information access systems is required to satisfy different information needs of users. For instance, e-commerce websites provide both search and recommendation functionalities to their users. 
Decades of research have been dedicated to developing specialized models for each information access scenario. For instance, lexical and semantic matching models developed for document retrieval in response to keyword queries are fundamentally different from the models used in state-of-the-art recommender systems. Inspired by Belkin and Croft's early examination of information retrieval and filtering systems \cite{Belkin:1992}, Zamani and Croft~\cite{JSR,JRL} hypothesized that joint modeling and optimization of search engines and collaborative filtering can lead to higher generalization and can improve both search and recommendation quality.
In this paper, we take this hypothesis even further \emph{towards} developing \textbf{universal information access}: 
\begin{quote}
    \emph{a unified model that can efficiently and effectively perform different information access functionalities}
\end{quote}
Successful implementation of universal information access could close the gap between the research communities working on various aspects of information access, such as the IR and RecSys communities. Potentially, all information access functionalities can benefit from universal modeling and optimization (i.e., transferring knowledge across information access functionalities). Moreover, such universal models can potentially reduce the engineering efforts required for developing and maintaining multiple information access systems separately. All of these points highlight the importance of research towards universal information access.

Thanks to the flexibility of dense retrieval models and their state-of-the-art performance in various information retrieval scenarios \cite{ANCE,DPR,Condenser}, we develop \framework--a novel dense retrieval framework for unified information access. \framework follows a bi-encoder architecture. The first encoder learns a dense vector for a given information access request (e.g., an item\footnote{In this paper, items refer to unstructured or semi-structured textual documents.} specified by the user or a search query submitted by a user) and an \textit{information access functionality} that the user is interested in (e.g., recommending complementary items or retrieving relevant items). Since a wide range of information access functionalities require personalization, \framework adjusts the obtained dense vectors based on the user's historical interactions by introducing a novel \emph{Attentive Personalization Network (APN)}. APN performs both content-based and collaborative personalization based on the user's past interactions with all the information access functionalities (e.g., user's past interactions with a search engine is also used for personalizing recommendation results). The second encoder learns a dense representation of the each information item in the collection. After training, these representations can be computed offline and an approximate nearest neighbor algorithm can be employed for efficient retrieval. Our proposed approach, the \framework model, utilizes a non-personalized pre-training and personalized fine-tuning strategy to improve representation learning. Additionally, we draw upon recent advancements in dense retrieval by incorporating the strategy of sampling hard negative items from various sources \cite{RocketQA,ANCE,ADORE}. By combining these training strategies, our approach aims to achieve better performance in various information access scenarios.

To train and evaluate the \framework framework, we focus on three information access functionalities. (1) \textbf{Keyword Search}: retrieving relevant items in response to a short textual query (e.g., retrieving `\texttt{iPhone 14 Pro}' for query `\texttt{apple smartphone}'), (2) \textbf{Query by Example}: retrieving items that are similar to a item specified by the user (e.g., retrieving `\texttt{Google Pixel 7 Pro}' as a similar item to `\texttt{iPhone 14 Pro}'), and (3) \textbf{Complementary Item Recommendation}: recommending information items that are complementary to a given item specified by the user (e.g., recommending `\texttt{AirPods Pro}' to a user who is interested in `\texttt{iPhone 14 Pro}'). \framework uses text description of each information access functionality, thus is \textit{generic} and extensible to a wide range of other information access functionalities.

We evaluate our model on a real-world dataset collected from user interactions with different information access systems on a major e-commerce website.\footnote{Even though our both datasets are from the e-commerce domain, the proposed approach is sufficiently generic to be applied to any domain. We are not aware of any other publicly available dataset beyond e-commerce that can be used to evaluate our task.} To improve reproducibility, we also extend our experiments to a dataset constructed from the Amazon ESCI data \cite{Reddy:2022:ESCI}, recently released as part of KDD Cup 2022.\footnote{KDD Cup 2022: \url{https://amazonkddcup.github.io/}} Extensive experiments demonstrate significant improvements compared to a wide range of competitive baselines for all three information access functionalities. We also demonstrate that up to 45\% NDCG@10 improvements can be obtained by jointly modeling all information access functionalities compared to their individual modeling using the same dense retrieval architecture. 

To summarize, the main contributions of this work include:
\begin{itemize}[leftmargin=*]
    \item Developing a generic and extensible dense retrieval framework that performs multiple information access functionalities.
    \item Demonstrating the feasibility of learning a single model that can perform a wide range of information access functionalities.
    \item Proposing an attentive personalized network with a two-stage training process and negative sampling strategies from various sources to improve personalized dense retrieval performance.
    \item Evaluating the proposed model on real-world data and demonstrating substantial gain compared to competitive baselines.
\end{itemize}

We open-source our implementation of the \framework framework to foster research towards developing universal information access.\footnote{\url{https://github.com/HansiZeng/UIA}}

\begin{table*}[t]
    \centering
    \caption{Properties of the three information access functionalities explored in this paper.}
    \vspace{-0.3cm}
    \resizebox{\textwidth}{!}{
    \begin{tabular}{lllll}\toprule
        & \texttt{Information Access Functionality} ($\mathcal{F}$) & \texttt{Information Access Request} ($\mathcal{R}$) & \texttt{User History} ($\mathcal{H}$) & \texttt{Item Information} ($\mathcal{I}$) \\\midrule
        1 & Keyword Search & A short keyword query & User's past queries and clicks & Content of candidate item \\
        2 & Query by Example & Content of an anchor item & User's past queries and clicks & Content of candidate item \\
        3 & Complementary Item Recommendation & Content of an anchor item & User's past queries and clicks & Content of candidate item\\\bottomrule
    \end{tabular}}
    \label{tab:properties}
    \vspace{-0.3cm}
\end{table*}

\section{Related Works}

\subsection{Personalized Information Access}
For one query, different users may have different intents or preferences. Hence, the ability of information access systems to capture the user's personalized preferences would play an important role in improving user experiences. For personalized search systems, there are several models have been proposed in recent years and have demonstrated search quality improvements compared to non-personalized models. For instance, HEM ~\cite{HEM} jointly models the user, query, and item representations using the doc2vec technique ~\cite{doc2vec} and ZAM~\cite{ZAM}  applies an attention-based method to capture the user dynamic information. 

In recommender systems, personalization plays an even bigger role. Early models utilized Collaborative Filtering (CF)~\cite{CF} to model the user's personalized preference. Recent studies \cite{text1,text2,image1,image2} employ deep neural networks to extract content information from items and seamlessly integrate this information with CF models.
However, CF models neglect the order of interactions in user history. To address this issue, models such as GRU4Rec~\cite{GRU4Rec}, SASRec~\cite{SASRec} and BERT4Rec~\cite{BERT4Rec} propose deep learning based sequential models to capture the user history preference for next-item prediction. 

\subsection{Dense Retrieval}
Accurate contextual representation of text using large-scale pre-trained language models has led to significant progress in various fields, including information retrieval. Combining these models with efficient approximate nearest neighbor search resulted in the development of dense retrieval models~\cite{ANCE,DPR}. These models fine-tune the pre-trained language models on the downstream information retrieval task~\cite{MSMARCO} and have shown significant performance improvement over strong lexical matching methods such as BM25 ~\cite{BM25}. One line of methods focuses on the optimization of dense retrieval models by producing better negative samples for contrastive loss. For example, DPR~\cite{DPR} uses the BM25 negatives as the source for hard-negative sampling and ANCE~\cite{ANCE} applies a self-sampling strategy for negative sampling. Models like RocketQA ~\cite{RocketQA} and Condenser ~\cite{Condenser} use large batch sizes and have demonstrated that it would be conducive to the stability of training. The other line of methods applies the knowledge distillation techniques that distill the knowledge from re-ranking models to train the dense retrieval models~\cite{LinInBatchNF,TAS-B,CL-DRD}. Adapting dense retrieval models to unseen data is also an active area of research \cite{BEIR}.



\subsection{Joint Search and Recommendation}
Recent results show that joint modeling of search and recommendation tasks can lead to better results compared to their individual training. JSR~\cite{JSR} is a framework that uses a task-specific layer on top of a shared representation learning network. It uses multi-task learning for optimization. More recently, SRJGraph~\cite{SRJGRaph} extends a similar approach by further applying graph convolution networks to capture higher-order interactions between users, queries, and items. 
These models are highly specialized for the search and recommendation tasks in hand and they are not simply extendable to other information access functionalities. On the other hand, we propose a dense retrieval model (\framework) that unifies different information access functionalities including search and recommendation. Unlike JSR and SRJGraph, \framework is able to jointly train different tasks without introducing additional parameters. \framework also contains a novel attentive personalization network (APN) that can capture the user's sequential interaction history and produces significantly better results compared to both JSR and SRJGraph.



\section{Methodology} 
\subsection{Task Formulation}
\label{sec:method:task}
Users can find and access information in a number of different ways. For instance, they can express their needs as a textual query and use a search engine to find relevant information. Alternatively, they can use the outcome of a recommender system without explicitly formulating their needs. This paper studies unified information access: \textit{developing and evaluating a unified model that can efficiently and effectively perform various information access functionalities}. Any information access model can be formulated as a scoring function that takes three input variables: 
\begin{enumerate}[leftmargin=*]
    \item \texttt{Information Access Request} ($\mathcal{R}$): it includes information about the current information access request, such as a search query, situational context (e.g., location and time) \cite{context_search}, and short-term context (e.g., session data) \cite{GRU4Rec,SASRec,BERT4Rec}. \texttt{Information Access Request} can be empty (i.e., zero-query retrieval) \cite{yang2016modelling}.
    
    \item \texttt{User History} ($\mathcal{H}$): it includes information about the user who issues $\mathcal{R}$, such as user's profile or their long-term interaction history.
    
    \item \texttt{Candidate Item Information} ($\mathcal{I}$): it includes the information about the candidate retrieval item, such as the item's content, author, and source.
\end{enumerate}

In order to develop a unified information access model, we introduce the \texttt{Information Access Functionality} ($\mathcal{F}$) as the fourth input variable. Without loss of generality, this paper focuses on the following three personalized information access functionalities:\footnote{Note that information access functionalities can go beyond these three functionalities, for example, by including multi-modal information (e.g., image queries and video items), contextual requests (e.g., sessions and conversations), and zero query retrieval or recommendation.}
\begin{enumerate}[leftmargin=*]
    \item \textbf{Keyword Search}: retrieving items in response to a short textual user query.
    \item \textbf{Query by Example}: retrieving items that are similar to a given item specified by a user.
    \item \textbf{Complementary Item Recommendation}: recommending complementary items to a given item specified by a user.
\end{enumerate}
The properties of each of these tasks with respect to the mentioned four input variables are listed in Table~\ref{tab:properties}. We intentionally include two information access functionalities that take identical inputs (i.e., Query by Example and Complementary Item Recommendation) but produce different outputs. They enable us to evaluate the model's ability in learning different information access functionalities.

Given the input variables introduced above, a unified information access model parameterized by $\theta$ for user $u$ at timestamp $t$ can be formally formulated as $f(\mathcal{F}^u_t, \mathcal{R}^u_t, \mathcal{H}^u_t, \mathcal{I}_i; \theta)$, where $\mathcal{F}^u_t$ is a textual description of the information access functionality being applied. For every \texttt{Information Access Request} ($\mathcal{R}^u_t$) at timestamp $t$, \texttt{User History} ($\mathcal{H}^u_t$) denotes a set of interactions that the user $u$ had prior to $t$. Therefore, $\mathcal{H}^u_t = \{(\mathcal{F}^u_1, \mathcal{R}^u_1, \mathcal{I}^u_1), (\mathcal{F}^u_2, \mathcal{R}^u_2,\mathcal{I}^u_2), \cdots, (\mathcal{F}^u_{t-1}, \mathcal{R}^u_{t-1}, \mathcal{I}^u_{t-1}) \}$ is a set of all the past $t-1$ interactions of the user $u$, each is represented by a triplet of the past user's request, the information access functionality that has been used, and the item that the user interacted with. For simplicity, in this paper, \texttt{Candidate Item Information} for the $i$\textsuperscript{th} item ($\mathcal{I}_i$) only includes the item's textual content. As mentioned in Table~\ref{tab:properties}, each $\mathcal{R}^u_t$ either represents a keyword query or the textual content of a given anchor item, depending on $\mathcal{F}^u_t$.

\begin{figure*}[t]
    \centering
    \vspace{-1cm}
    \includegraphics[width=0.8\textwidth]{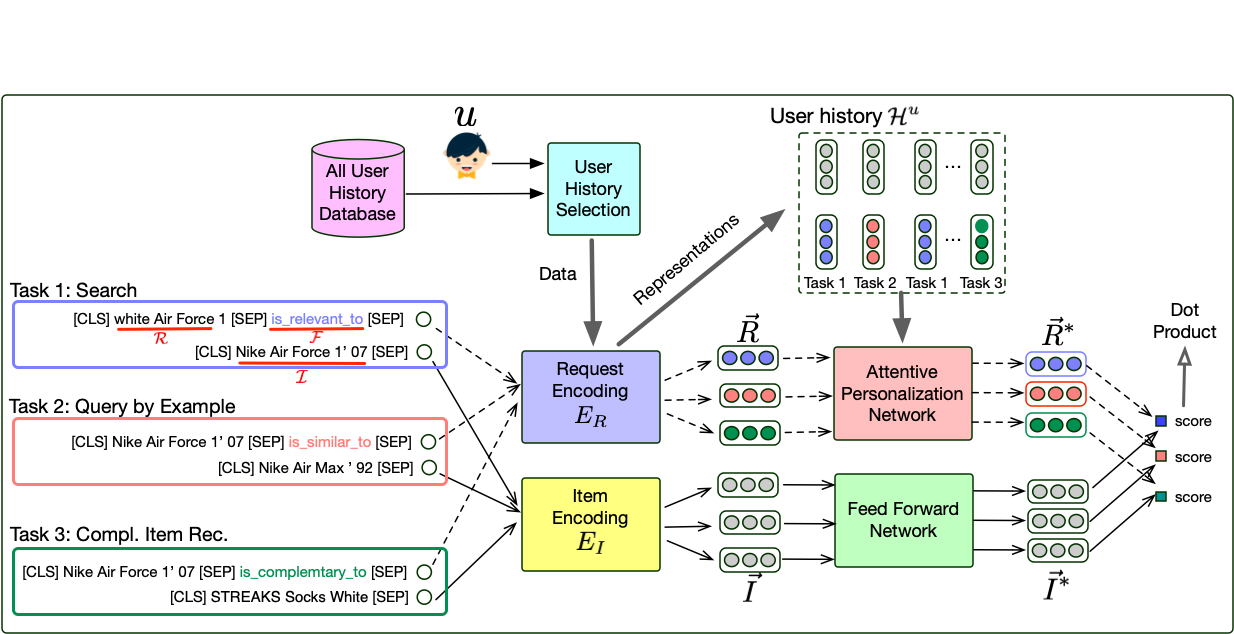}
    \vspace{-0.2cm}
    \caption{A high-level overview of the \framework framework.}
    \label{fig:arch}
    \vspace{-0.3cm}
\end{figure*}

\vspace{-0.3cm}
\subsection{Overview of the \framework Framework}
This paper proposes \framework, a dense retrieval framework for unified information access. A high-level overview of the \framework framework is depicted in Figure~\ref{fig:arch}. \framework follows a bi-encoder architecture. For each user $u$ at timestamp $t$, the first encoder takes \texttt{Information Access Functionality} ($\mathcal{F}^u_t$) and \texttt{Request} ($\mathcal{R}^u_t$) and produces a dense representation of the request. It then uses a novel \emph{Attentive Personalization Network} to further enhance the request representation through both collaborative and content-based personalization using the user's historical interaction data. The second encoder takes the content of a candidate item and produces a latent dense representation, which is fed to a feed-forward network in order to adjust the representations based on the personalized request vector. \framework uses the inner product to compute the similarity of encoded request and item. Therefore, both encoders should have the same output dimensionality. In the following subsections, we present the implementation and optimization of \framework. 

\subsection{\framework Architecture}
As demonstrated in Figure~\ref{fig:arch}, \framework consists of four major components: (1) Request Encoding, (2) Item Encoding, (3) User History Selection \& Encoding, and (4) Attentive Personalization Network. 

\paragraph{\textbf{Request Encoding}} To implement the request encoder, we use a pre-trained large language model (denoted by $\mathbf{E}$) that often feeds a subword tokenization of the input to an embedding layer followed by a number of Transformer layers and produces a dense vector representation for each input token. In this paper, we use BERT-base \cite{BERT} to encode each information access functionality ($\mathcal{F}^u_t$) and request ($\mathcal{R}^u_t$), as follows:
$$\vec{R}^u_t = \mathbf{E}_\mathcal{R}(\texttt{[CLS]}~\mathcal{R}^u_t~\texttt{[SEP]}~\mathcal{F}^u_t~ \texttt{[SEP]})$$

An example of the Request Encoding input is shown in Figure~\ref{fig:arch}.

\paragraph{\textbf{Candidate Item Encoding}} Similar to the request encoder, this component also uses a pre-trained BERT-base model for representing each candidate item $\mathcal{I}_i$. It represents the item's information (i.e., content) by encoding the following input:
$$\vec{I}_i = \mathbf{E}_\mathcal{I}(\texttt{[CLS]}~\mathcal{I}_i~\texttt{[SEP]})$$

For both the request and candidate item encoders, we use the \texttt{[CLS]} token representation as the encoder's output.

\paragraph{\textbf{User History Selection and Encoding}} For content-based personalization of the request encoding $\vec{R}^u_t$, this paper simply uses the last $N$ interactions of the user, i.e., $\{(\mathcal{F}^u_{t-N}, \mathcal{R}^u_{t-N}, \mathcal{I}^u_{t-N}), \cdots, (\mathcal{F}^u_{t-1}, \mathcal{R}^u_{t-1}, \mathcal{I}^u_{t-1}) \}$. We produce two encoding vectors for each of the user's past interactions: one for the past information access functionality and request ($\mathbf{E}_\mathcal{R}(\texttt{[CLS]}~\mathcal{R}^u_{t'}~\texttt{[SEP]}~\mathcal{F}^u_{t'}~ \texttt{[SEP]}) : \forall t-N \leq t' \leq t-1$) and another for the item that the user interacted with ($\mathbf{E}_\mathcal{I}(\texttt{[CLS]}~\mathcal{I}^u_{t'}~\texttt{[SEP]}) : \forall t-N \leq t' \leq t-1$). \framework uses the same $\mathbf{E}_\mathcal{R}$ and $\mathbf{E}_\mathcal{I}$ encoders to ensure that the user history encodings are represented in the same space as the model's input for timestamp $t$. All parameters in \framework, including the encoder models, are trained \emph{end-to-end}, and thus they get updated during training. 

Therefore, the encoded history would be a set of $2 \times N$ vectors as follows: $\{(\vec{R}^u_{t-N}, \vec{I}^u_{t-N}), (\vec{R}^u_{t-N+1}, \vec{I}^u_{t-N+1}), \cdots, (\vec{R}^u_{t-1}, \vec{I}^u_{t-1})\}$

\paragraph{\textbf{Attentive Personalized Network}}
To produce personalized representation of $\vec{R}^u_{t}$, we propose a novel Attentive Personalization Network (APN) that performs both content-based and collaborative personalization. For content-based personalization, it learns attention weights from encodings of the user's past $N$ interactions to the current request. Note that for each information access functionality, it is able to use the user's past interactions with all functionalities (e.g., both search and recommendation). For collaborative personalization, it learns a latent representation for each user and information access functionality based on all the past interactions. 

\begin{figure}[t]
    \centering
    \includegraphics[width=0.9\linewidth]{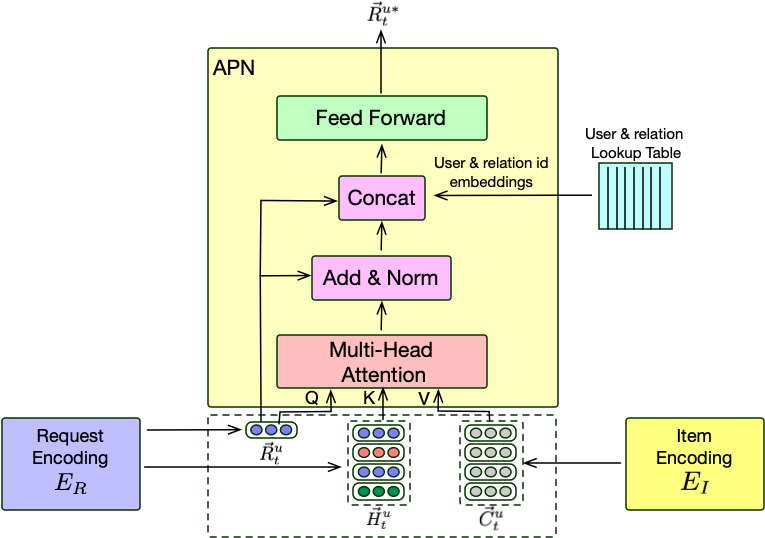}
    \caption{The Attentive Personalization Network (APN).}
    \label{fig:transformer-layer}
    \vspace{-0.5cm}
\end{figure}

A high-level overview of APN architecture is presented in Figure~\ref{fig:transformer-layer}.
At each timestamp $t$, APN takes the request encoding $\vec{R}_t$ and the last $N$ user history encodings $\{(\vec{R}^u_{t-N}, \vec{I}^u_{t-N}), \cdots, (\vec{R}^u_{t-1}, \vec{I}^u_{t-1})\}$. It converts the user history encodings to two matrices: (1) $H^u_t \in \mathbb{R}^{N \times d}$ whose rows are equal to the past request encodings, and (2) $C^u_t \in \mathbb{R}^{N \times d}$ whose rows are equal to the past interacted item (clicked) encodings.

APN consists of $N_h$ attention heads, where the $j$\textsuperscript{th} attention function contains three parameter matrices $\theta_j^Q \in \mathbb{R}^{d \times l}$, $\theta_j^K \in \mathbb{R}^{d \times l}$, and $\theta_j^V \in \mathbb{R}^{d \times l_v}$. It learns attention weights from the user's interaction history to their current request. Therefore, each APN layer computes the following three vectors:
\begin{align}
    {Q_j} = \vec{R}^u_t . \theta_j^Q, \quad
    {K_j} = H^u_t . \theta_j^K, \quad
    {V_j} = C^u_t . \theta_j^V \nonumber
\end{align}

Therefore, $Q_j \in \mathbb{R}^{1 \times l}$, $K_j \in \mathbb{R}^{N \times l}$, and $V_j \in \mathbb{R}^{N \times l_v}$. Based on these three matrices, APN uses the following attention mechanism that is similar to Transformer \cite{Attentionallyounned}:
\begin{equation}
    \text{Attn}(Q_j, K_j, V_j) = \text{softmax} ( \frac{Q_j K_j^T}{\sqrt{l}}) V_j \nonumber
\end{equation}

The output of all attention functions are then concatenated: $\text{concat}(\{\text{Attn}(Q_j, K_j, V_j) \}_{j=1}^{N_h}) \in \mathbb{R}^{1 \times N_{h} l_{v}}$. Following Transformer architecture and inspired by residual and layer norm in neural networks, we feed this matrix to an Add \& Norm layer.

For \textbf{collaborative personalization}, the Attentive Personalization Network also learns a user embedding matrix $\textbf{E}_\mathcal{U} \in \mathbb{R}^{|\mathcal{U}| \times l_u}$, where $|\mathcal{U}|$ denotes the number of users and $l_u$ is the user embedding dimensionality. For each user $u$, we select the associated user embedding vector $\vec{u}$ from $\textbf{E}_\mathcal{U}$ (i.e., user embedding lookup). To distinguish between the user behavior dealing with different information access functionalities, we also learn an embedding matrix for information access functionalities $\textbf{E}_\mathcal{F} \in \mathbb{R}^{|\mathcal{F}| \times l_f}$, where $|\mathcal{F}|$ denotes the number of information access functionalities (e.g., three in our case) and $l_f$ is the functionality embedding dimensionality. At timestamp $t$, we select the associated functionality embedding vector $\vec{f}$ from $\textbf{E}_\mathcal{F}$.

We later concatenate the output of Add \& Norm layer with $\vec{u}$ and $\vec{f}$ and feed this vector to a feed-forward layer with non-linear activation (ReLU). This produces a personalized representation of the current request, denoted by $\vec{R}^{u*}_t$. 

Since the personalization component is only applied to the request representation, we use a feed-forward network for the candidate item vectors to adjust their representations with the new personalized semantic space and obtain $\vec{I}^*_i$ (See Figure~\ref{fig:arch}).

\vspace{-0.4cm}
\subsection{\framework Optimization}
We propose a two stage optimization process: (1) non-personalized pre-training, and (2) personalized fine-tuning. The reason is that real-world systems often deals with a large number of new and cold-start users with no or limited historical interactions. Thus, personalization cannot help, yet we can use their data for non-personalized pre-training.
\vspace{-0.3cm}
\paragraph{\textbf{Non-Personalized Pre-Training}}
We construct a non-personalized training set by aggregating the training data across all users. For each training instance $k$ is in the form of $(\mathcal{F}_k, \mathcal{R}_k, \mathcal{I}_k, \mathcal{Y}_k)$ where $\mathcal{Y}_k$ is the ground truth label. We get the output vector of the Request Encoder and Candidate Item Encoder, i.e., $\vec{R}_k$ and $\vec{I}_k$, respectively. We compute the non-personalized matching of the request and the candidate item using dot product: $\vec{R}_k \cdot \vec{I}_k$.

The training data only contains positive instances (i.e., user interactions), thus appropriate negative sampling is required. We apply a two-phase negative sampling and training strategy. 

\textit{Phase 1:} for each request in training data, we randomly sample negative items from the top 200 items   retrieved by BM25. We set the ratio of negative samples to positive training instances to 1. We then train the model using a cross entropy loss function. Note that in addition to BM25 negative, we also use in-batch negatives.

\textit{Phase 2:} Once the model is trained in Phase 1, we use the trained Candidate Item Encoder $\mathbf{E}_{\mathcal{I}}$ to encode all items in the collection and create an approximate nearest neighbor (ANN) index using the Faiss library \cite{faiss}. The constructed index is then used to retrieve items for each request in the training data and random negatives are sampled from the top 200 retrieved documents. This self-negative sampling strategy has successfully been used in a number of dense retrieval models, such as ANCE \cite{ANCE} and RANCE \cite{RANCE}. We set the ratio of negative samples to positive training instances to 1. Similar to Phase 1, we re-train the model using a cross-entropy loss function that also uses in-batch negatives.

\vspace{-0.2cm}
\paragraph{\textbf{Personalized Fine-Tuning}}
In non-personalized pre-training, only the parameters of $\textbf{E}_{\mathcal{R}}$ and $\textbf{E}_{\mathcal{I}}$ are adjusted. We then add the personalization part of the framework and re-create the training data to include the user information and their past interactions. We then use the personalized representation of each request and candidate item in the training data, i.e., $\vec{R}^*_k$ and $\vec{I^*_k}$ respectively (see Figure~\ref{fig:arch}). Then, we use dot product to compute their matching score: $\vec{R}^*_k \cdot \vec{I^*_k}$.

For negative sampling, we use BM25 results in addition to in-batch negatives (similar to Phase 1 in non-personalized pre-training). We use cross entropy loss function for training.

\begin{table}[t]
    \caption{Statistics of the datasets constructed for this study.}
    \vspace{-0.3cm}
    \resizebox{\linewidth}{!}{
    \begin{tabular}{l!{\color{lightgray}\vrule}ll!}
    \toprule
     \textbf{Property} & \textbf{Lowe's Data} & \textbf{Amazon ESCI}  \\
     \midrule
     \# unique users & 893,619 & - \\
     \# unique queries & 953,773 & 68,139 \\
     \# items in the collection & 2,260,878 & 1,216,070 \\
     \# interactions  for keyword search & 4,075,996 & 874,087 \\
     \# interactions  for query by example & 968,778 & 2,254,779 \\
     \# interactions  for complementary item rec. & 329,992 & 303,481 \\
     \bottomrule
    \end{tabular}}
    \vspace{-0.3cm}
    \label{tab:statistics}
\end{table}

\label{personalized learniing}

\begin{table*}[t]
    \caption{Experimental results on the Lowe's dataset. ``-'' means the model is not applicable to the search task. Superscripts $^\triangle$ and $^\blacktriangle$  refer to significant improvements compared to all task-specific training baselines and all baselines (including joint training), respectively (p\_value < 0.01).}
    \vspace{-0.3cm}
    \centering
    \begin{tabular}{p{3.5cm}!{\color{lightgray}\vrule}lll!{\color{lightgray}\vrule}lll!{\color{lightgray}\vrule}lll!} 
    \toprule
        & \multicolumn{3}{c!{\color{lightgray}\vrule}}{\textbf{Keyword Search}} & \multicolumn{3}{c!{\color{lightgray}\vrule}}{\textbf{Query by Example}} & \multicolumn{3}{c!}{\textbf{Complementary Item Rec.}}\\
         \textbf{Model} & MRR & NDCG & Recall & MRR & NDCG & Recall & MRR & NDCG & Recall \\
         \midrule
    BM25 & 0.089 & 0.095 & 0.367 & 0.153 & 0.167 & 0.584 & 0.016 & 0.014 & 0.111 \\
    \midrule
    \multicolumn{4}{l!}{\textbf{Task-Specific Training}} \\
    NCF & - & - & - & 0.132 & 0.147 & 0.351 & 0.117 &  0.118 & 0.236   \\
    DPR & 0.188 & 0.192 & 0.578 & 0.171 & 0.180 & 0.598 & 0.153 & 0.156 & 0.487\\
    ANCE  & 0.193 & 0.199 & 0.582 & 0.176 & 0.188 & 0.601 & 0.159 & 0.158 & 0.494 \\
    RocketQA & 0.201 & 0.207 & 0.595 & 0.189 & 0.204 & 0.613 & 0.174 & 0.176 & 0.507 \\
    Context-Aware DPR &  0.324 & 0.377 & 0.848 & 0.311 & 0.356 & 0.860 & 0.278 & 0.283 & 0.707  \\
    Context-Aware ANCE& 0.332 & 0.385 & 0.856 & 0.317 & 0.361 & 0.866 & 0.289 & 0.292 & 0.714   \\
    Context-Aware RocketQA  & 0.335 & 0.389 & 0.861 & 0.326 & 0.369 & 0.874 & 0.300 & 0.304 & 0.723 \\
    SASRec++ & - & - & - & 0.305 & 0.347 & 0.836 & 0.271 & 0.264 & 0.695  \\ 
    BERT4Rec++ & - & - & - & 0.314 & 0.354 & 0.851 & 0.283 & 0.279 & 0.703  \\
    \midrule
    \multicolumn{4}{l!}{\textbf{Joint Training}} \\
    JSR & 0.324 & 0.379 & 0.853 & 0.349 & 0.380 & 0.878 & 0.325 & 0.317 & 0.760 \\
    JSR+BERT4Rec++ & 0.337 & 0.394 & 0.871 & 0.415 & 0.479 & 0.919 & 0.421 & 0.419 & 0.820  \\
    SRJGraph & 0.336 & 0.392 & 0.874 & 0.416 & 0.478 & 0.921  & 0.423 & 0.420 & 0.822 \\
    \framework & \textbf{0.340}$^\triangle$ & \textbf{0.399}$^\blacktriangle$ & \textbf{0.880}$^\blacktriangle$ &  \textbf{0.433}$^\blacktriangle$ & \textbf{0.495}$^\blacktriangle$ & \textbf{0.945}$^\blacktriangle$ & \textbf{0.438}$^\blacktriangle$ & \textbf{0.432}$^\blacktriangle$ & \textbf{0.836}$^\blacktriangle$    \\
    \bottomrule
    \end{tabular}
    \label{tab:results:lowes}
    \vspace{-0.3cm}
\end{table*}

\section{Experiments}
\subsection{Experiment Settings}
\subsubsection{\textbf{Dataset}}
Evaluating the proposed model is challenging, since there is no public dataset that provide large-scale training data for multiple information access functionalities. Therefore, we made significant efforts to create two datasets. The first one is a private real-world data obtained from the Lowe's website--the second-largest hardware chain in the world. The second dataset is built from the public Amazon ESCI data \cite{Reddy:2022:ESCI} recently released as part of KDD Cup 2022. Table~\ref{tab:statistics} summarizes the statistics of the datasets. 

The \textbf{Lowe's dataset} is a large-scale e-commerce data containing over 5.3 million user interactions for over obtained from over 890K  unique users. The item collection in this data includes over 2.2 million products. The Lowe's dataset contains three types of user-item interaction data: (1) Keyword Search (i.e., query-item click data); (2) Query by Example (i.e., item-item click data for similar items); and (3) Complementary Item Recommendation (i.e., item-item click data for complementary item pairs). This dataset also includes the anonymized user ID and timestamp associated with each interaction. 

The \textbf{Amazon ESCI dataset} \cite{Reddy:2022:ESCI} is adopted from KDD Cup 2022 - Task 2 that contains 3 tasks related to product information access. We modify the data to fit our experimental setting. 
Each data point in KDD Cup 2022 - Task 2 is a (query, item, label) triplet. The label is one of the following four classes: (1) Exact (E) means that the item is relevant to (an exact match for) the query; (2) Substitute (S) means that the item is related to the query but not an exact match (partially relevant); (3) Complement (C) means the item is not relevant to the query but can complement the relevant item with label E; and (4) Irrelevant (I). 
Let $I_E(q)$, $I_S(q)$, and $I_C(q)$ respectively denote a set of all items with a label E, label S, and label C for query $q \in Q$. We used the following procedure to construct our three datasets: (1) Keyword Search: $\{(q, i): \forall q \in Q \land i \in I_E(q)\}$, (2) Query by Example: $\{(i_1, i_2): \forall q \in Q \land i_1 \in I_E(q) \land i_2 \in I_S(q)\}$, and (3) Complementary Item Recommendation: $\{(i_1, i_2): \forall q \in Q \land i_1 \in I_E(q) \land i_2 \in I_C(q)\}$. 



Note that the {Amazon ESCI dataset} does not contain user identifiers, hence no personalization is performed for this dataset. To the best of our knowledge, there is no public dataset with user identifiers that includes multiple information access functionalities.



\subsubsection{\textbf{Evaluation Protocols}}
For the Lowe's dataset, we use a leave-last-out data splitting strategy, which has been widely used in the literature~\cite{AttentiveCF,He2018NAISNA,Kabbur2013FISMFI}. For each of the three information access functionalities, we use the user's most recent interaction for testing, their second most recent interaction for validation, and the rest for training. This results in over 890K interactions in each of the test and validation sets. This realistic data splitting strategy enables us to evaluate the model's ability based on the future interactions that the user may have with the system. Since there is no user identifiers or timestamp available in the Amazon ESCI dataset, we randomly select $80\%$ of requests (e.g., query text or query item) for training, $10\%$ for validation, and the remaining $10\%$ for testing.

To evaluate the models, we report performance in terms of a wide range of metrics: MRR@10 (MRR for short), NDCG@10 (NDCG for short), and Recall@50 (Recall for short). We use two-tailed paired t-test with Bonferroni correction to identify statistically significant improvements (p\_value < 0.01).


\begin{table*}[t]
    \centering
    \caption{Experimental results on the Amazon ESCI dataset. Superscript $^\blacktriangle$  refers to significant improvements compared to all baselines (p\_value < 0.01).}
    \vspace{-0.3cm}
    \begin{tabular}{p{3cm}!{\color{lightgray}\vrule}lll!
    {\color{lightgray}\vrule}lll!
    {\color{lightgray}\vrule}lll!}
    \toprule
     & \multicolumn{3}{c!}{\textbf{Keyword Search}} & \multicolumn{3}{c!}{\textbf{Query by Example}} & \multicolumn{3}{c!}{\textbf{Complementary Item Rec.}} \\
     \textbf{Model} & MRR & NDCG & Recall & MRR & NDCG & Recall@50 & MRR & NDCG & Recall \\
    \midrule
    BM25 & 0.513  &  0.351 & 0.494 & 0.017 & 0.011 & 0.084 & 0.030 & 0.032 & 0.165  \\
    \midrule
    \multicolumn{4}{l!}{\textbf{Task-Specific Training}} \\
    DPR  & 0.505 & 0.347  & 0.511 & 0.235 & 0.174  & 0.527 & 0.434 & 0.450  & 0.838 \\
    ANCE &  0.522 & 0.354  & 0.519 & 0.237 & 0.178  & 0.531 & 0.431 & 0.443 & 0.825 \\
    RocketQA & 0.526 & 0.357 & 0.525 & 0.244 & 0.185 &  0.538 & 0.445 & 0.458 & 0.847  \\
    \midrule
        \multicolumn{4}{l!}{\textbf{Joint Training}} \\
    JSR & 0.528 & 0.355 & 0.527  & 0.243 & 0.192 & 0.536 & 0.477 & 0.484 & 0.853  \\
    SRJGraph  &  0.526 & 0.351 & 0.522 & 0.241 & 0.187 & 0.540 & 0.479 & 0.488 & 0.855 \\
    \framework  & \textbf{0.532}$^\blacktriangle$ & \textbf{0.360}$^\blacktriangle$ & \textbf{0.533}$^\blacktriangle$ &  \textbf{0.251}$^\blacktriangle$ & \textbf{0.199}$^\blacktriangle$ & \textbf{0.543}$^\blacktriangle$&  \textbf{0.490}$^\blacktriangle$ &  \textbf{0.493}$^\blacktriangle$  & \textbf{0.868}$^\blacktriangle$  \\
    \bottomrule
    \end{tabular}
    \label{tab:results:esci}
    \vspace{-0.3cm}
\end{table*}

\subsubsection{\textbf{Implementation Details}}
We use BERT-base \cite{BERT} available on HuggingFace~\cite{huggingface} as the pre-trained language model in our models. For the Lowe's dataset, the pre-trained weights is loaded from the checkpoint \footnote{\url{https://huggingface.co/bert-base-uncased}} and for Amazon ESCI dataset (which is smaller), the pre-trained weights are obtained from the checkpoint \footnote{\url{https://huggingface.co/sentence-transformers/msmarco-bert-base-dot-v5}}. 
We use the model's performance on the validation set in terms of NDCG to select the hyper-parameters. 

We chose the number of training epoch from $[8, 12, 16, 24, 48]$. We set the user's historical interactions ($N$) to 5. The batch size is empirically set to $384$. The learning rate for pre-training and personalized fine-tuning is empirically set to $7e^{-6}$ and $7e^{-5}$, respectively. For personalized fine-tuning, we only keep users with at least $10$ interactions in search and query by example tasks, and at least $5$ interactions in the complementary item recommendation task. The hidden dimension $d$ is 768, number of heads $N_h$ is $12$, hidden dimension of key and value in each head are $l = l_v = 64$. The dimension of user embedding is $l_u = 128$ and of functionality embedding is $l_f=64$. We use Adam \cite{Adam} as the optimizer. 



\subsubsection{\textbf{Baselines}}
We use a wide range of baselines in our experiments, from term matching models to dense retrieval models. We also include collaborative and sequential recommendation baselines, when appropriate. The baselines are listed below:

\begin{itemize}[leftmargin=*]
    \item \textbf{BM25} \cite{BM25}: This is a simple yet effective bag-of-word retrieval model that uses query term frequency, inverse document frequency, and document length to compute relevance scores. 
    \item \textbf{NCF} \cite{NCF}: This is an effective collaborative filtering model, which combines generalized matrix factorization and a multi-layer perceptron approach for recommendation. It only learns from item-item interactions and cannot be applied to keyword search tasks.
    \item \textbf{DPR} \cite{DPR}: DPR is a dense retrieval model that samples negative documents from the items retrieved by BM25 in addition to in-batch negative sampling. DPR only uses the last request (query text or query item) and does not perform personalization.
    \item \textbf{Context-Aware DPR}: We extend the DPR model to include user history (personalization). To this aim, we simply concatenate the current user request with their past interactions, separated by a \texttt{[SEP]} token and feed it to the query encoder. 
    \item \textbf{ANCE} \cite{ANCE}: ANCE is an effective dense retrieval model that uses the model itself to mine hard negative samples. Similar to DPR, ANCE is not capable of personalization, so we also include a \textbf{Context-Aware ANCE} using a similar approach used for Context-Aware DPR.
    \item \textbf{RocketQA} \cite{RocketQA}: RocketQA is a state-of-the-art dense retrieval model. It utilizes the large batch size and denoised negative samples for more robust contrastive learning. Similar to DPR and ANCE, RocketQA is not capable of personalization, so we also include a \textbf{Context-Aware RocketQA} for user sequential modeling.  
    \item \textbf{BERT4Rec++}: BERT4Rec~\cite{BERT4Rec} is a sequential recommendation model that represents the user interaction history using BERT for predicting the next item. The original BERT4Rec model takes item IDs and predict the next item ID in the sequence. We improve BERT4Rec by encoding item content too. We use BERT for content embedding. We call this approach BERT4Rec++.
    \item \textbf{SASRec++}: SASRec \cite{SASRec} is a sequential recommendation model uses the self-attention mechanism to identify which items are ``relevant'' to the user interaction history for the next item prediction. It cannot be used for the keyword search tasks. Similar to the last baseline, we use BERT for content embedding and call it as model SASRec++.
    \item \textbf{JSR} \cite{JSR}: This is a neural framework that jointly learns the search and recommendation tasks. Each task has a task-specific layer over the base shared network.  
    \item \textbf{JSR + BERT4Rec++}: The original JSR uses the user ID to encode user information. We improve the JSR performance by using the representation from BERT4Rec++ to encode the user content.
    \item \textbf{SRJGraph} \cite{SRJGRaph}: This is a recent framework based on neural graph convolution that jointly models the search and recommendation tasks. 
 \end{itemize}  

 We use the provided public code to implement BM25, NCF, DPR, ANCE, RocketQA, SASRec++ and BERT4Rec. We implemented JSR and SRJGraph which don’t have public implementations. Based on how the training data is used, we classify the training of models into two categories: (1) task-specific training; (2) joint training. Task-specific training means that the baseline is only trained on the target task (i.e. for evaluation on the search task, they are trained on the search data only). Joint training baselines can access all tasks' data as UIA does. We use the same BERT backbone with the same initial pre-trained weights to train all dense retrieval baselines (DPR, ANCE, RocketQA, and ours). All context-aware variants consume historial data as their input in reverse chronological order. For SASRec++ and BERT4Rec++, the number of additional transformer layers is chosen from [1,2,4] and each transformer layer contains 12 heads and each head's hidden dimension is 64. The task-specific layer for JSR and SRJGraph is a single dense layer with hidden dimension 768 and we initiated their share networks by the same BERT model as dense retrieval baselines. The learning rate for all baselines is chosen from [1e-4, 7e-5, 1e-5, 7e-6] based on dev set results.


\begin{table}[t]
    \centering
    \caption{Ablation study results on the Lowe's dataset in terms of NDCG for keyword search (Search), query by example (QBE), and complementary item recommendation (CIR). Superscripts $^\triangledown$ denote significantly lower performance compared to \framework (p\_value < 0.01).}
    \vspace{-0.3cm}
    \resizebox{\linewidth}{!}{
    \begin{tabular}{ll!{\color{lightgray}\vrule}ccc!}
    \toprule
     && \textbf{Search} & \textbf{QBE} & \textbf{CIR} \\
    \midrule
    - &\framework  & {0.399} & 0.495 & {0.432}\\\midrule
    1 & w/o encoding $\mathcal{F}$ & 0.371$^\triangledown$  & 0.347$^\triangledown$  & 0.284$^\triangledown$  \\
    2 & w/o joint optimization & 0.391$^\triangledown$  & 0.369$^\triangledown$  & 0.298$^\triangledown$  \\ 
    3 & w/o APN & 0.207$^\triangledown$  & 0.214$^\triangledown$  &  0.176$^\triangledown$  \\
    4 & w/o \emph{combined} content-based personalization & 0.397 & 0.482$^\triangledown$  & 0.411$^\triangledown$  \\ 
    5 & w/o collaborative personalization & 0.378$^\triangledown$  & {0.507} & 0.419$^\triangledown$  \\
    

    \bottomrule
    \end{tabular}
    }
    \label{tab:ablation study}
    \vspace{-0.3cm}
\end{table}
\subsection{Experimental Results}
\subsubsection{\textbf{Main Results}}
The performance of \framework and the baselines on the Lowe's dataset for all the three information access functionalities is reported in Table~\ref{tab:results:lowes}. The results demonstrate that BM25 and NCF perform poorly compared to the deep learning based models. Note that NCF cannot be used for the keyword search task, as it is a collaborative filtering approach. We also observe that context-aware variation of dense retrieval models substantially outperform the original DPR and ANCE models. This demonstrates the importance of personalization for information access in e-commerce. Besides, BERT4Rec++ has better results than SASRec++, which implies that BERT that is a stack of multiple Transformer layers is capable of better capturing user history behaviors. Among models that are trained on a corresponding single task, Context-Aware ANCE achieves the best performance. 

Table~\ref{tab:results:lowes} also shows that joint training models perform better than task-specific models, especially for query by example and complementary item recommendation tasks that have substantially smaller training data compared to the keyword search task. All in all, \framework outperforms all the baselines. The improvements are statistically significant in nearly all cases, except for MRR in Keyword Search which is only significant compared to the task-specific models. 

We extend our experiments to the Amazon ESCI dataset and report the result in Table~\ref{tab:results:esci}. Note that ESCI does not contain user identifiers, thus the baselines that cannot perform without personalization are omitted from Table~\ref{tab:results:esci}. In this dataset, complementary item recommendation benefit the most from joint training. The reason is due to the size of the dataset for this information access functionality. It is smaller than the other two functionalities, thus it benefit the most from knowledge transfer across tasks.
These results suggest that \framework again performs better than all the baselines. The improvements are generally smaller than the Lowe's dataset and it is due to the lack of personalization in \framework for this dataset.  

We find a concurrent work \cite{Su2022OneEA} which aims to create a universal encoder for all retrieval tasks. We utilized their publicly available pre-trained checkpoint for zero-shot evaluation on the Amazon ESCI dataset, however, its performance was inferior to BM25. As an example, the MRR@10 in the complementary recommendation task was 0.223 compared to 0.230 by BM25. We surmise that this disparity could be due to the method being trained on datasets that prioritize semantic text similarity \cite{Conneau2018SentEvalAE,MSMARCO} and overlooks complementary or subsititue item matching signals.

\vspace{-1em}
\subsubsection{\textbf{Ablation Study}}
To demonstrate the impact of each novel component used in \framework, we conduct a thorough ablation study on the Lowe's dataset that contain temporal information as well as anonymized user identifiers. For the sake of space, we only report NDCG@10 values in Table~\ref{tab:ablation study}. We make the following observations:

From Row 1, encoding the information access functionality (i.e., $\mathcal{F}$) is found crucial. The reason is that the model does not know what items are expected to be retrieved given the input request. Thus, it behaves similarly across all information access functionalities. The performance gain in query by example and complementary item recommendation is substantially higher than in keyword search. There are two reasons: (1) the search data in Lowe's dataset is at least 300\% larger than each of the other information access functionalities (see Table~\ref{tab:statistics}); and (2) the input to both query by example and complementary item recommendation are identical 

From Row 2, we can conclude that all three information access functionalities benefit from joint optimization. The improvement observed in keyword search task is marginal, but 34\% and 45\% relative improvements are observed in query by example and complementary item recommendation, respectively. Note that Lowe's dataset contains fewer data points for these two tasks and thus they can substantially benefit from joint optimization. 

Row 3 shows that personalization (both content-based and collaborative) using APN has a significant impact on the models performance. Using APN in \framework leads to 93\%, 72\%, and 145\% NDCG@10 improvements in keyword search, query by example, and complementary item recommendation, respectively. These improvements are mostly coming from the content-based personalization. That being said, in the following, we show that APN also benefits from using the combination of all interaction history as well as collaborative personalization.

Row 4 reports the results for \framework where for each information access functionality, only the associated user history is used for personalization. For example, only past search interactions of users are used for the search functionality, as opposed to \framework that uses all the past interactions combined. From this result, we observe that \framework benefits from including historical user interactions from all three information access functionalities. Again, query by example and complementary item recommendation benefit the most, due to their smaller training data size (and thus fewer historical user interactions) compared to the search task. 

Finally, Row 5 demonstrates the results for \framework without collaborative personalization, i.e., without user and relation embedding in APN's final layer (see Figure~\ref{fig:transformer-layer}). We show that adding user and relation embedding does not improve the model performance consistently across information access functionalities. Collaborative personalization is found to be helpful in keyword search and complementary item recommendation but not in query by example. The possible explanation might be that the structural information provided by user-relation-item ids is conducive to identify the item-item complementary and query-item relevance relations, but content information which is important in query by example (finding similar items)  can be better distilled from content-based personalization without using the user embeddings.

\begin{figure}
    \centering
    \includegraphics[width=\linewidth]{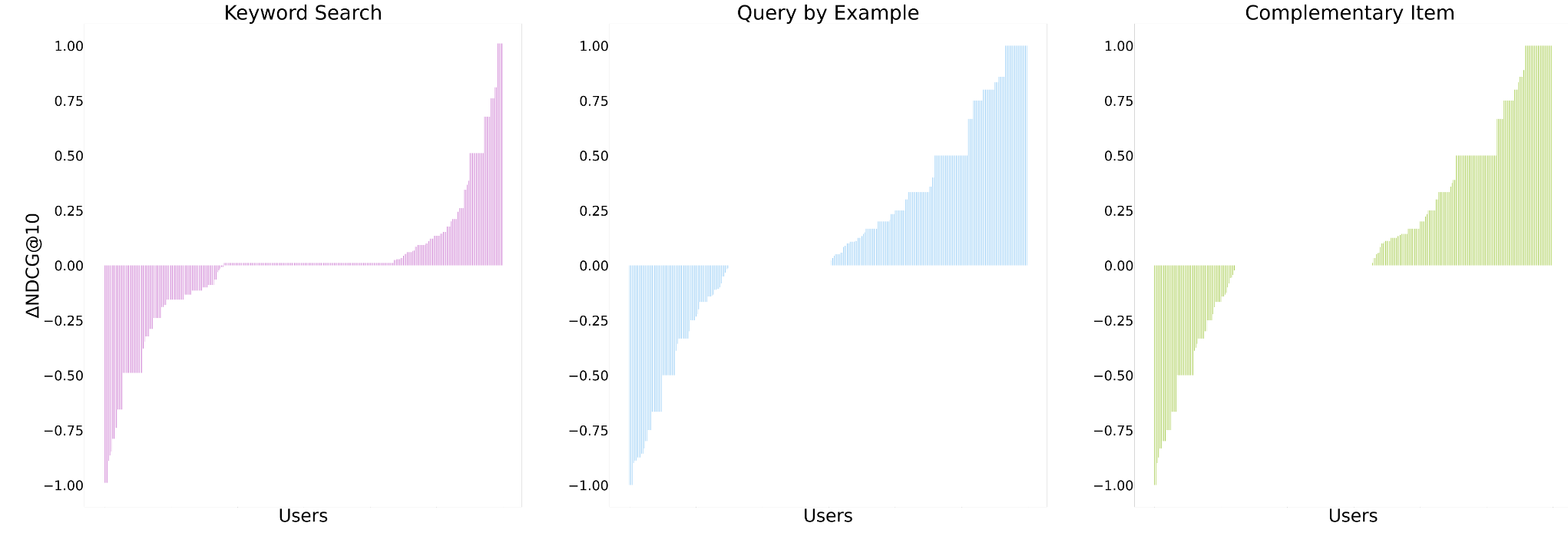}
    \caption{$\Delta$NDCG (sorted) at user level between \framework and TS-\framework (i.e., \framework without joint training) on the Lowe's dataset.}
    \label{fig:score diff.}
    \vspace{-0.3cm}
\end{figure}

\begin{figure}
    \centering
    \includegraphics[width=\linewidth]{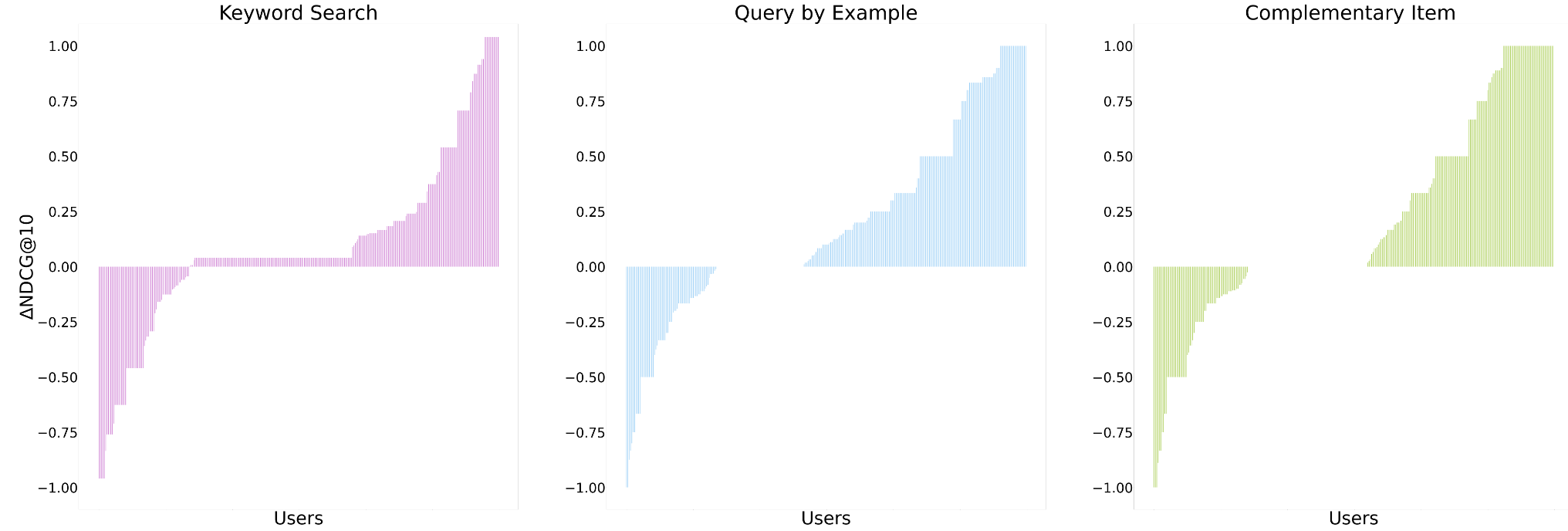}
    \caption{$\Delta$NDCG (sorted) at user level between \framework and \framework without APN on the Lowe's dataset.}
    \label{fig:personalized_score_diff.}
    \vspace{-0.3cm}
\end{figure}

\begin{table*}[t]
     \caption{\small The top 1 retrieved item between \framework and TS-\framework. The item is represented as the concatenation of title and its category separated by ";". \textcolor{green}{\ymark} means the retrieved item is relevant and \textcolor{red}{\xmark} means irrelevant.}
     \centering
     \resizebox{\linewidth}{!}{
     \begin{tabular}{p{65mm}!{\color{lightgray}\vrule}p{65mm}!{\color{lightgray}\vrule}p{65mm}!
     {\color{lightgray}\vrule}p{30mm}!}
     \toprule 
     \textbf{Query or query\_item} & \textbf{1st Ranked Item by \framework} & \textbf{1st Ranked Item  by TS-\framework} & \textbf{Error Reason} \\
    \midrule
     \textbf{Keyword Search}: \newline 6x6 privacy fence & Freedom Brighton 6-ft H x 6-ft W White Vinyl Flat-top Fence Panel ; Vinyl Fencing \textcolor{green}{\ymark}  & HOFT Solutions HOFT Kit A6- One 73 in. End Post Black and Hardware  ; Outdoor Privacy Screens \textcolor{red}{\xmark}  & Only part of apsect matching  \\
    \midrule
     \textbf{Keyword Search}: \newline ev charger  & Westinghouse 40-Volt Charger Lithium Ion (li-ion) ; Cordless Power Equipment Batteries \& Chargers \textcolor{red}{\xmark}  & LectronTo NEMA 14-50 Plug J1772 Cable EV Charger Level 2 40-Amp Freestanding Single Electric Car Charger ; Electric Car Chargers \textcolor{green}{\ymark}  & Only part of apsect matching \\
    \midrule 
     \textbf{Query by Example}: \newline
     Arrow 10-ft W x 20-ft L x 8-ft H Eggshell Metal Carport; Carports & Arrow 10-ft x 15.27-ft Eggshell Metal Carport ; Carports \textcolor{green}{\ymark}  & Arrow 10-ft W x 20-ft L x 8-ft H Eggshell Metal Carport  ; Carports \textcolor{red}{\xmark} & Exact the same item as query\_item. \\
    \midrule
     \textbf{Compl. Item Rec.}: \newline
     Owens Corning Oakridge 32.8-sq ft Driftwood Laminated Architectural Roof Shingles ; Roof Shingles & Owens Corning VentSure 15-in x 48-in Black Plastic Stick Roof Ridge Vent ; Roof Ridge Vents \textcolor{green}{\ymark}  & Owens Corning DecoRidge 20-lin ft Driftwood Hip and Ridge Roof Shingles ; Roof Shingles \textcolor{red}{\xmark}  & Similar but not complementary item \\
     \bottomrule
     \end{tabular}}
     \label{tab:case study}
     \vspace{-0.3cm}
 \end{table*}

\subsubsection{\textbf{The Impact of Joint Training}}
Not all users benefit from joint training, since many of them may only use only one of the information access functionalities. 
To further investigate the effect of joint modeling, we compare the results between \framework and its task-specific (or single task) variation (i.e., the same model only trained on the target information access functionality, denoted by TS-\framework) at user level. We compute the performance of models in terms of NDCG for each user in the test set. The performance difference between \framework and TS-\framework for users is sorted and plotted in Figure~\ref{fig:score diff.}. According to the graph, 60+\% of users benefit from joint optimization in query by example and complementary item recommendation tasks. We can observe that $\Delta$NDCG for 5-10\% of users in both of these tasks is equal to 1. Meaning that for these users joint optimization produces perfect ranking while task-specific training does not retrieve or recommend any relevant item. This is while we do not observe a significant number of users with $\Delta$NDCG=-1. In the keyword search task, the plot is more balanced. The reason is that the training data for this task is substantially larger than the other two tasks and thus it does not benefit a lot from joint optimization. These three plots also explains why the observed improvements for query by example and complementary item recommendation in Table~\ref{tab:results:lowes} are larger than those for the keyword search task. Given these three plots, we suggest future work to focus on ensemble of task-specific and joint optimization.



\subsubsection{\textbf{The Impact of Personalization}}
To better understand the impact of personalization in \framework, we compare the results obtained from \framework and its non-personalized variation (i.e., \framework without APN). The performance difference (i.e., $\Delta$NDCG) at the user level (sorted) is plotted in Figure~\ref{fig:personalized_score_diff.} for all three information access functionalities. According to the figure, not all users benefit from personalization. This is expected especially for new and cold-start users and also for situations where user's information need is different from their past interactions. The plots show that personalization has a negative impact on less than a quarter of users and a larger set of users take advantage of personalization. These plots also show that personalization has a larger impact on query by example and complementary item recommendation compared to keyword search. This is expected, as in keyword search, the user's information need is often clear from the query, while it is more difficult to infer user's need in query by example and complementary item recommendation. Predicting the need for personalization in future work can substantially improve \framework.

\vspace{-0.2cm}
\subsubsection{\textbf{Case Study}} Our case study, which compares the top-1 retrieved item by \framework and its task-specific training counterpart TS-\framework, is presented in Table \ref{tab:case study}. The results show that both models struggle with partial attribute matching errors in the keyword search task. For example, when searching for ``6x6 privacy fence,'' the top-1 item returned by TS-\framework is ``outdoor privacy screens'' which only matches the ``privacy'' attribute but misses the ``fence'' attribute. Similarly, \framework returns an irrelevant item for the query ``ev charger.'' In the query by example task, the top-1 retrieved item by the task-specific training counterpart (TS-\framework) is often the same as the query item. This may be due to high term overlap in the positive training pairs, resulting in over-fitting of the model. Joint modeling with other tasks can partially mitigate this issue as different tasks serve as regularizers. In the complementary item recommendation task, \framework demonstrates superiority over TS-\framework by accurately identifying complementary items, such as correctly matching ``roof ridge vents'' with ``roof shingle'' despite the weak term matching signal.
\vspace{-0.3cm}

\section{Conclusions and Future Directions}
This paper studied the feasibility of implementing universal information access systems and proposed a generic and extensible dense retrieval framework, called \framework, for this purpose. \framework encodes each information access functionality in addition to the user's request and takes advantage of approximate nearest neighbor search for efficient retrieval and recommendation of items. In addition, it introduces a novel attentive personalization network to extend the application of \framework to personalized information access tasks. 
We constructed two datasets to evaluate our approach, one large-scale real-world private data and another data built from the recent Amazon ESCI dataset. We demonstrated that \framework significantly outperforms competitive baselines and conducted extensive empirical exploration and ablation study to evaluate various aspects of the model.
Based on the results reported in this paper, we envision a bright future (both short- and long-term) for universal information access. In the future, we will extend this work to other information access functionalities, such as question answering. We are also interested in evaluating the performance of the models beyond e-commerce applications. 
Therefore, future work will focus on data creation, organizing evaluation campaigns, and exploring ensemble approaches involving task-specific and universal information access models while predicting the need for personalization.

\vspace{-0.2cm }
\section{Acknowledgments}
This work was supported in part by the Center for Intelligent Information Retrieval, in part by NSF grant number 2143434, in part by Lowes, and in part by the Office of Naval Research contract number N000142212688. Any opinions, findings and conclusions or recommendations expressed in this material are those of the authors and do not necessarily reflect those of the sponsor.

\bibliographystyle{ACM-Reference-Format}
\bibliography{sample-base}

\appendix

\end{document}